\documentclass[conference]{IEEEtran}

\usepackage[
    colorlinks=true,
	linkcolor=black,
	citecolor=black,
    urlcolor=blue
]{hyperref}

\usepackage{cite}
\usepackage{amsmath,amssymb,amsfonts}
\usepackage{algorithm}
\usepackage{algorithmic}
\usepackage{array}
\usepackage{graphicx}
\usepackage{textcomp}
\usepackage{xcolor}
\usepackage{hyperref}
\usepackage[english]{babel}
\usepackage{orcidlink}
\usepackage{booktabs}
\usepackage{placeins} 
\usepackage{tikz}
\usepackage{float}

\newcolumntype{C}[1]{>{\centering\arraybackslash}p{#1}}

\usetikzlibrary{positioning,shapes.geometric} 

\def\BibTeX{{\rm B\kern-.05em{\sc i\kern-.025em b}\kern-.08em
    T\kern-.1667em\lower.7ex\hbox{E}\kern-.125emX}}

\begin{document}

\title{Multi-Agent Framework Leveraging Knowledge Graphs for Virtual Commissioning Models}

\author{
	\IEEEauthorblockN{ Max Diekmann\orcidlink{0009-0004-9696-0264}\IEEEauthorrefmark{1}\IEEEauthorrefmark{2}\IEEEauthorrefmark{3}, Jonas Nitzler\orcidlink{0000-0002-7258-5664}\IEEEauthorrefmark{1}, Jan Fischer\IEEEauthorrefmark{1}, Hans-J\"urgen Pfisterer\IEEEauthorrefmark{4}, Dirk Hartmann\orcidlink{0000-0002-6020-1061}\IEEEauthorrefmark{1}\IEEEauthorrefmark{2} }
	\IEEEauthorblockA{\IEEEauthorrefmark{1} Siemens AG, Munich, Germany}
	\IEEEauthorblockA{\IEEEauthorrefmark{2} Technical University of Darmstadt, Darmstadt, Germany}
	\IEEEauthorblockA{\IEEEauthorrefmark{4} University of Applied Sciences Osnabr\"uck, Osnabr\"uck, Germany}
	\IEEEauthorblockA{\IEEEauthorrefmark{3} Corresponding author: max.diekmann@siemens.com} }
\maketitle

\begin{abstract} Virtual commissioning models (VCMs) of discrete manufacturing systems are
	used to validate automation behavior before physical deployment, but creating and
	maintaining them remains labor-intensive. Relevant engineering information is distributed
	across programmable logic controller (PLC) engineering projects such as Siemens TIA
	Portal and kinematic simulation models such as Siemens NX Mechatronics Concept Designer
	(NX MCD), where it is stored in tool-specific, incompatible data structures. In practice,
	block-structured IEC 61131-3-based PLC programs and variables are engineered separately
	from rigid-body and kinematic simulation objects such as parts, joints, sensors, and
	actuators. As a result, understanding system behavior, generating simulation components,
	and mapping PLC variables to corresponding control-relevant simulation objects require
	cross-domain expertise and remain largely manual. This paper presents a
	knowledge-graph-grounded multi-agent framework for semi-automated support of VCM
	development. A deterministic setup process systematically extracts structured data from
	Siemens TIA Portal and Siemens NX MCD and transforms both sources into graph-based
	representations in a shared graph database. The framework utilizes a hierarchical
	multi-agent architecture to support three selected task classes in early-stage VCM
	development: system understanding, simulation component generation, and cross-domain
	signal mapping. For these tasks, the framework provides grounded natural-language access
	to engineering knowledge, template-guided generation of executable NX Open journal
	scripts, and ranked mapping suggestions between PLC variables and NX MCD simulation
	objects. The approach was evaluated on a laboratory-scale discrete manufacturing system.
	The results demonstrate that transforming engineering data into an agent-accessible and
	interaction-ready representation can reduce manual cross-domain interpretation effort and
	make recurring VCM engineering tasks more directly actionable. The contribution of this
	work lies in integrating structured extraction, graph-based knowledge representation, and
	agent-based task execution into a unified assistance workflow for semi-automated VCM
	development.
\end{abstract}


\section{Introduction}~\label{sec:introduction} Virtual commissioning models (VCMs) of discrete manufacturing
systems~\cite{verein_deutscher_ingenieure_virtuelle_2025} are an established means of validating
automation software before physical deployment and thereby reducing commissioning effort,
risk, and iteration time in manufacturing systems\cite{hoffmann_virtual_2010,striffler_concepts_2023}. In discrete manufacturing, VCMs
combine control logic, mechanical structure, and simulation-relevant behavior in an
offline environment that enables engineers to test sequences, signals, and machine
interactions before commissioning the real system. In this paper, a VCM is understood as
an offline, pre-deployment virtual representation for commissioning support and is
therefore distinct from a runtime digital twin, which requires ongoing synchronization
with a physical asset~\cite{hartmann_real-time_2021,barricelli_survey_2019}. Despite their practical relevance,
the development of VCMs remains labor-intensive because the required engineering
knowledge is split across programmable logic controller (PLC) engineering environments
and kinematic simulation tools, each with tool-specific control and simulation data
schemas~\cite{striffler_concepts_2023}. A central technical challenge is integrating information
across two disconnected engineering environments: automation logic development and
simulation modeling~\cite{hartmann_digital_2025}. In industrial practice, PLC software is created using
block-structured programs under IEC 61131-3~\cite{international_electrotechnical_commission_programmable_2025}, whereas rigid body
simulation models are built in tools such as Siemens NX Mechatronics Concept Designer (NX
MCD) using object-oriented component hierarchies. As a result, information relevant for
VCM development is fragmented across software environments. Expert engineers must
therefore manually interpret PLC programs and variables, trace their usage, identify
corresponding control-relevant simulation objects, and create simulation components such
as sensors and actuators across tool boundaries. This process requires expert-level
cross-domain knowledge~\cite{yang_leveraging_2025} and becomes particularly demanding in early development phases
or for users who are not yet familiar with the system structure~\cite{striffler_concepts_2023}.

Recent research offers foundational building blocks for addressing this problem, but not
yet an integrated solution for the present use case. Knowledge graphs have demonstrated
potential for structuring and linking heterogeneous industrial information and for
enhancing intelligent digital representations of technical systems~\cite{sahlab_knowledge_2021}. At the same time, large language models (LLMs) and
multi-agent systems are increasingly explored as methods to support engineering and
digital-twin-related workflows through natural language interaction, retrieval support,
and task decomposition~\cite{yang_leveraging_2025,greis_multi_agent_2025}. However, existing approaches typically
focus either on semantic data integration or on artificial intelligence (AI)-assisted
engineering support in isolation. This paper addresses the gap between combining PLC
engineering data and simulation data by proposing a knowledge-graph-grounded multi-agent
framework for semi-automated support of VCM development in discrete manufacturing. The
framework first applies a deterministic extraction (Phase I - setup) and integration
process to structured data from Siemens TIA Portal and Siemens NX MCD and then uses a
hierarchical multi-agent architecture (Phase II interaction) on top of the resulting
graph-based representation layer to support three task classes: system understanding,
simulation component generation, and cross-domain signal mapping. More specifically, the
framework provides (1) grounded natural-language access to engineering knowledge, (2)
template-guided generation of NX Open journal scripts for simulation components, and (3)
ranked mapping suggestions between PLC variables and NX MCD objects.

We selected these three task classes because they represent three recurring requirements
in early stage VCM development. In this phase, relevant engineering information is
distributed across PLC engineering data and simulation model data, while the latter is
often available only as static 3D computer-aided-design (CAD) geometry rather than as a
configured simulation model. System understanding therefore forms a necessary starting
point for identifying relevant program structures, variables, and mechanical components.
On that basis, simulation components can be generated in a targeted manner from the
available geometric representation. Once automation, - and simulation elements are
available in a usable form, cross-domain signal mapping becomes a further essential task.
The selected task classes were therefore intended to cover a representative sequence of
early-stage VCM activities rather than the full VCM workflow.

The main contributions of this work are threefold. First, we present a structured
extraction and integration pipeline that transforms heterogeneous PLC and kinematic
engineering data into a graph-based knowledge representation. Second, we introduce a
multi-agent framework that leverages retrieval, reasoning, and component generation for
these three task classes. Third, we evaluate the approach on a laboratory-scale discrete
manufacturing system and demonstrate its feasibility across all three tasks. The
remainder of this paper is structured as follows. \autoref{sec:related_work} reviews related work on virtual commissioning-related
frameworks, knowledge graphs, and agent-based engineering support. \autoref{sec:methodology} describes the methodology. \autoref{sec:results} presents the results. \autoref{sec:conclusion} concludes the findings, limitations, and implications for future
research and engineering practice.
\section{Related Work}~\label{sec:related_work} Research relevant to this paper lies in existing virtual
commissioning approaches, semantic industrial data integration, and AI-supported
engineering assistance. Virtual commissioning is widely recognized as an effective means
of validating automation behavior before physical deployment and of reducing
commissioning effort in manufacturing systems~\cite{hoffmann_virtual_2010}. At the same time, prior reviews identify kinematic
simulation model creation as a major obstacle to broader industrial adoption, since the
generation still requires substantial effort and domain expertise~\cite{striffler_concepts_2023}. Existing automatic and semi-automatic simulation
generation approaches are typically tied to specific data sources, modelling aspects, or
toolchains~\cite{striffler_concepts_2023}. Closely related work on digital twins highlights the
broader relevance of executable and simulation-based virtual representations across the
lifecycle of industrial systems~\cite{barricelli_survey_2019,singh_digital_2021}, while application-oriented studies on 3D
models and Siemens NX MCD demonstrate the practical value of simulation environments for
manufacturing processes and control validation~\cite{li_design_2024,zheng_towards_2021}. However, these approaches generally assume that
relevant simulation structures or engineering assets are already available in usable
form, leaving open how PLC - and simulation- engineering knowledge can be systematically
integrated for VCM development. A second relevant strand concerns the integration of
heterogeneous engineering knowledge. Knowledge graphs have been proposed as an effective
representation for entities, relations, and dependencies across engineering domains while
supporting semantic querying and reasoning~\cite{sahlab_knowledge_2021,zheng_towards_2021,zhu_knowledge_2021}. Standardized
approaches such as AutomationML and the Asset Administration Shell target interoperable
engineering data exchange and digital asset representation~\cite{drath_automationml_2021,international_electrotechnical_commission_asset_2023}. In
contrast, this work focuses on direct extraction from proprietary tool instances and
their transformation into a graph-based representation tailored to concrete VCM tasks.
Prior work has also shown that knowledge graphs can support tasks such as simulation
reuse, synchronization, and automatic kinematic model generation~\cite{listl_architecture_2023}. More recent studies further explore the use of LLMs to
construct or enrich digital-twin-related knowledge graphs in domain-specific settings~\cite{yang_llm-based_2024,mandal_llmasmmkg_2024}. Still, most of these approaches focus on
semantic integration and knowledge management rather than on VCM-specific support for
simulation component generation or cross-domain signal mapping.

A third strand of related work addresses retrieval-augmented generation (RAG),
graph-based retrieval (graphRAG), and multi-agent systems for digital-twin-related tasks.
RAG has emerged as an effective strategy for grounding LLM outputs in external knowledge~\cite{shuster_retrieval_2021,lewis_retrieval-augmented_2020,zhao_retrieval-augmented_2026}.
Recent work on graphRAG shows that knowledge graphs improve retrieval and support
reasoning over entity-relation data~\cite{zhu_graph-based_2026,edge_local_2024}. In parallel, multi-agent systems are
increasingly used to decompose digital-twin-related workflows into specialized reasoning,
coordination, and interaction functions~\cite{greis_multi_agent_2025,marah_madtwin_2024}. Additional studies show that LLM-driven
digital twin architectures can integrate multi-modal data, capture temporal system
characteristics, and maintain traceability in application-specific settings~\cite{sun_empowering_2024}. Recent PLC-focused agentic LLM systems further indicate
growing interest in AI support for industrial engineering, but they primarily focus on
engineering assistance within TIA Portal e.g. on PLC code generation and verification~\cite{liu_agents4plc_2026,adnyana_benchmarking_2026}.

Recent advances in digital-twin-related AI increasingly address retrieval, orchestration,
and system intelligence, but largely remain at the architectural level rather than
targeting virtual commissioning engineering assistance specifically. This paper addresses
the gap by extracting structured knowledge from Siemens TIA Portal and Siemens NX MCD,
integrating it into a knowledge graph database, and employing a multi-agent framework to
support system understanding, simulation component generation, and cross-domain signal
mapping in semi-automated VCM workflows.

\section{Methodology}~\label{sec:methodology} This section provides an overview of the methodology, the utilized
data sources and their analysis, followed by the applied data extraction pipelines to the
automatic knowledge graph creation. Data retrieval pipeline and the multi-agent
architecture are described in the last two subsections, respectively.

\subsection{Study Design and System Overview}
\begin{figure}[htbp]
	\centering
	\begin{tikzpicture}
		\tikzset{myNode/.style={anchor=center,inner sep=1pt,fill=white,fill opacity=0.70,rounded
					corners=2pt}}
		\node[anchor=south west,inner sep=0] (image) at (0,0) {\includegraphics[width=\columnwidth]{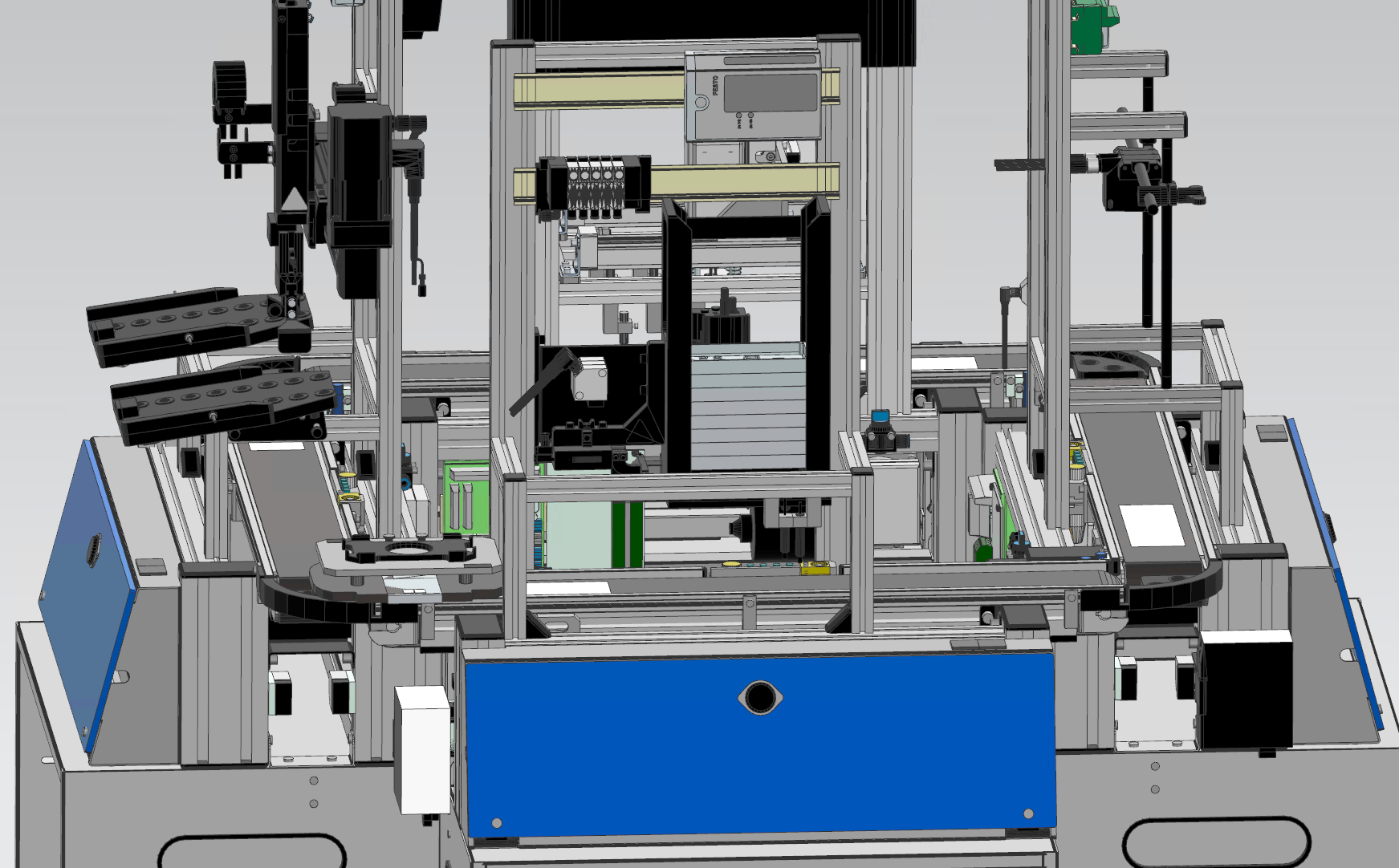}};
		\begin{scope}[x={(image.south east)},y={(image.north west)}]
			\node[anchor=south west,inner sep=0] (subimage) at (0.01,0.01) {\includegraphics[width=0.09\columnwidth]{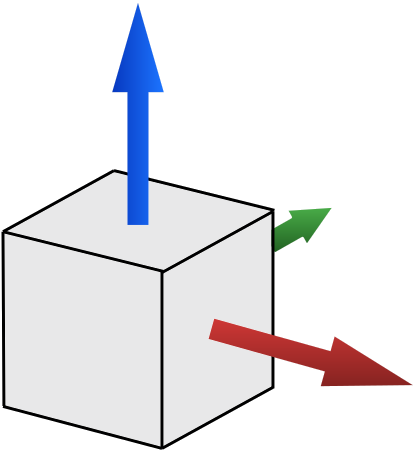}};

			\draw (.04,.19) node[myNode] {\textcolor{blue!80!white}{\textbf{Z}}};
			\draw (.1,.115) node[myNode] {\textcolor{green!65!black}{\textbf{Y}}};
			\draw (.115,.03) node[myNode] {\textcolor{red!90!black}{\textbf{X}}};
		\end{scope}
	\end{tikzpicture}
	\caption{Overview of the manufacturing under study}~\label{fig:magazine_station}
\end{figure}
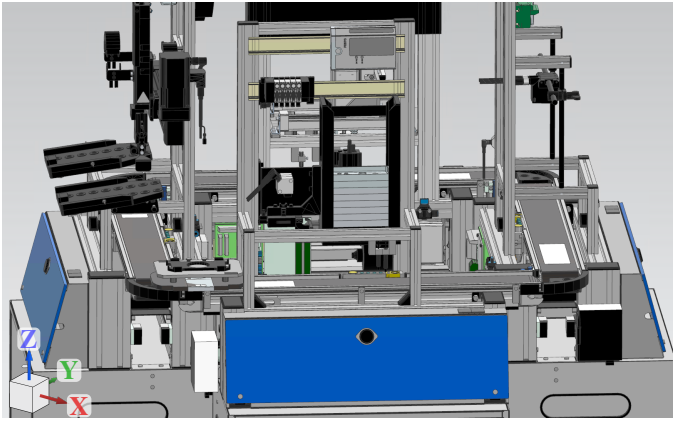

The manufacturing system under study is the Festo Didactic CP Factory~\cite{festo_festo_nodate}, a modular, laboratory-scale production platform in which
workpiece carriers transport an assembly along a conveyor-based line through a series of
application stations. To ground the evaluation queries in the physical system, \autoref{fig:magazine_station} shows the first station module. It contains a stacking
shaft holding front or back shells, a lifting cylinder that lowers to deposit a shell
onto a passing workpiece carrier, a separation cylinder that releases one shell at a
time, and a stopper cylinder that halts the incoming carrier. Detection is handled by
sensors signaling magazine-empty state, carrier presence, and shell type. These physical
roles are what the queries in \autoref{tab:eval_queries} probe across the three use cases.

\begin{figure}[htb]
	\centering
	\resizebox{\columnwidth}{!}{%
		\begin{tikzpicture}

			\node[align=center] at (-5.35,0.0) {inputs};
			\node[draw, fill=black!6, rounded corners, align=center, minimum width=2.6cm, minimum
				height=0.8cm] (in_tia) at (-2.0,0.0) {TIA Portal\\project};
			\node[draw, fill=black!6, rounded corners, align=center, minimum width=2.6cm, minimum
				height=0.8cm] (in_nx) at (2.0,0.0) {NX MCD\\kinematic model};

			\node[align=center] at (-5.35,-1.45) {phase~I:\\setup};
			\node[draw, fill=black!6, rounded corners, align=center, minimum width=2.6cm, minimum
				height=0.9cm] (ext_tia) at (-2.0,-1.45) {deterministic PLC\\logic extraction};
			\node[draw, fill=black!6, rounded corners, align=center, minimum width=2.6cm, minimum
				height=0.9cm] (ext_nx) at (2.0,-1.45) {deterministic NX MCD\\kinematic extraction};

			\node[align=center] at (-5.35,-3.05) {phase~II:\\interaction};
			\node[draw, fill=black!6, rounded corners, align=center, minimum width=6.1cm, minimum
				height=1.1cm] (core) at (0,-3.05) {knowledge graphs + graphRAG +\\ three-tier multi agent system};

			\node[align=center] at (-5.35,-4.8) {outputs};
			\node[draw, fill=black!6, rounded corners, align=center, minimum width=2.0cm, minimum
				height=0.9cm] (a1) at (-2.8,-4.8) {grounded natural-\\language response};
			\node[draw, fill=black!6, rounded corners, align=center, minimum width=2.0cm, minimum
				height=0.9cm] (a2) at (0,-4.8) {NX Open\\journal};
			\node[draw, fill=black!6, rounded corners, align=center, minimum width=2.0cm, minimum
				height=0.9cm] (a3) at (3.0,-4.8) {ranked PLC-to-NX\\mapping suggestions};

			\node[draw, fill=black!10, rounded corners, align=center, minimum width=6.1cm, minimum
				height=0.8cm] (hitl) at (0,-6.0) {human-in-the-loop: final engineering acceptance before
				deployment};

			\draw[->, thick] (in_tia) -- (ext_tia);
			\draw[->, thick] (in_nx) -- (ext_nx);
			\draw[->, thick] (ext_tia) -- (core);
			\draw[->, thick] (ext_nx) -- (core);
			\draw[->, thick] (core) -- (a1);
			\draw[->, thick] (core) -- (a2);
			\draw[->, thick] (core) -- (a3);
			\draw[->, thick] (a2) -- (hitl);
		\end{tikzpicture} }
	\caption{Two-phase methodology pipeline. Phase~I (setup) runs deterministic extraction and
		graph preparation scripts from Siemens TIA Portal and NX MCD. Phase~II (interaction) runs
		the multi-agent integration and produces three outputs: grounded natural-language
		responses, NX Open journals, and ranked PLC-to-NX mapping suggestions.}\label{fig:methodology_pipeline}
\end{figure}
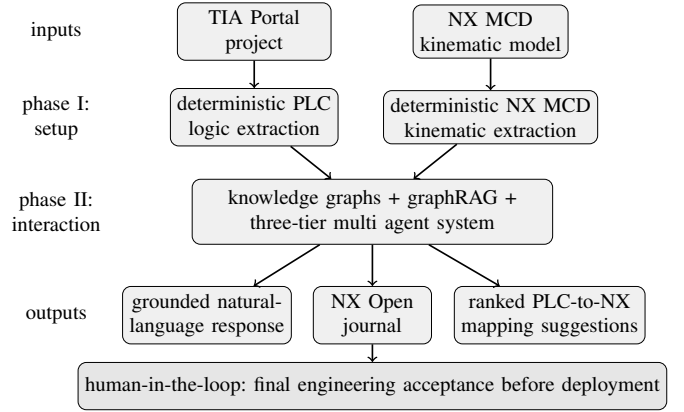

This work is a design and validation study of our framework for supporting VCM. To
separate deterministic preprocessing from multi-agent implementation, the workflow is
organized into two phases, see~\autoref{fig:methodology_pipeline}. Phase I (setup phase) executes deterministic
extraction and graph-construction scripts once per revision, see~\autoref{subsec:tia_extract} - \autoref{subsec:graph_creation}. Phase II (interaction phase) addresses the multi-agent
integration on top of the prepared graph representation, see~\autoref{subsec:agent_orch}.

Our framework boundary comprises two inputs and three task classes. Input 1 is a Siemens
TIA Portal V19 project containing PLC control logic. Input 2 is a Siemens NX MCD
kinematic model, which may be empty, partially configured, or populated with mechanical
components, joints, sensors, and control interfaces. We extract structured data from both
sources and construct two knowledge graphs that preserve PLC interconnections and the NX
MCD hierarchy. On this basis, we implement a graphRAG-enabled multi-agent system to
support the three evaluated task classes.

\FloatBarrier%

\subsection{PLC Logic Extraction from Siemens TIA Portal}~\label{subsec:tia_extract} This subsection describes deterministic preprocessing that
prepares PLC metadata for the later multi-agent interaction. The extraction process
targets PLC projects implemented in IEC 61131-3 compliant languages, with particular
focus on the block-oriented program structures commonly used in TIA Portal. The relevant
logic is contained in organization blocks (OBs), function blocks (FBs), and functions
(FCs). In addition to block-level information, the extraction pipeline captures input and
output variables, internal variables, and call relations between networks. For compact
notation in the following algorithm, let $P$ denote the TIA Portal project, $X$ the
exported XML representation, $T$ the global PLC tags, $D$ the data blocks, $B$ the logic
blocks (FB/FC/OB), and $N$ the extracted network collection.

\autoref{alg:tia_extraction} then summarizes the extraction workflow as $P \rightarrow X \rightarrow T, D, B \rightarrow N \rightarrow (T, D, B, N)$: the project $P$ is exported to XML $X$, from which $T$, $D$,
$B$ are extracted; $N$ is then initialized and populated by parsing each block network
into its constituent elements (header, access, parts, and wires), after which ($T$, $D$,
$B$, $N$) is serialized for knowledge graph ingestion.

\begin{algorithm}[htb]
	\caption{TIA Portal data extraction pipeline}~\label{alg:tia_extraction}
	\footnotesize
	\begin{algorithmic}[1]
		\renewcommand{\algorithmicrequire}{\textbf{Input:}}
		\renewcommand{\algorithmicensure}{\textbf{Output:}}
		\REQUIRE TIA Portal project file \(P\)
		\ENSURE Extracted TIA variables \(T, D\), blocks \(B\), and networks \(N\)

		\vspace{0.5em}
		\textit{initialize extraction} :
		\vspace{0.2em}
		\STATE \(X \gets \text{OpennessAPI.ExportToXML}(P)\)
		\STATE \(T \gets \text{ExtractPLCTags}(X)\)
		\STATE \(D \gets \text{ExtractDataBlocks}(X)\)
		\STATE \(B \gets \text{ExtractLogicBlocks}(X)\)
		\STATE \(N \gets \emptyset\)

		\vspace{0.2em}
		\textit{loop process}
		\vspace{0.2em}
		\FOR{each block \(b \in B\)}
		\STATE \(componentData.block \gets \text{ExtractBlockMetadata}(b)\)
		\STATE \(compileUnits \gets \text{ExtractCompileUnits}(b)\)
		\FOR{each unit \(cu \in compileUnits\)}
		\STATE \(header \gets \text{ExtractNetworkHeader}(cu)\)
		\STATE \(access \gets \text{ExtractNetworkAccess}(cu)\)
		\STATE \(parts \gets \text{ExtractNetworkParts}(cu)\)
		\STATE \(wires \gets \text{ExtractNetworkWires}(cu)\)
		\STATE \(network \gets \text{CombineNetworkElements}(header,\ access,\)
		\STATE \hspace{1.5em}\(parts,\ wires)\)
		\STATE \(componentData.networks.append(network)\)
		\ENDFOR
		\STATE \(N \gets N \cup \{componentData\}\)
		\ENDFOR

		\STATE \textbf{return} \(T, D, B, N\)
	\end{algorithmic}
\end{algorithm}

Language-specific parsing is then applied per block, see
\autoref{alg:tia_extraction_language}. Following IEC 61131-3 terminology, LAD denotes
Ladder Diagram, FBD Function Block Diagram, SCL Structured Control Language, and STL
Statement List. LAD/FBD are processed as graphical networks by traversing parts (logic
element within a network) and wires (connecting logic elements within a network). SCL/STL
are retained as original source code in a dedicated code attribute rather than deeply
parsed at this stage. This preserves structural and semantic integrity for downstream
agent-based interpretation. Nested calls are resolved recursively across FB/FC boundaries
and stored as explicit call dependencies.

For each PLC block, the extraction pipeline stores block identifier, block type,
parent-child relation, interface variables, referenced variables, and call dependencies.
For each variable, the pipeline stores symbolic name, datatype, comment, and block
affiliation (e.g. input, output or DB variable). The return of the metadata is a
structured JSON format.

\begin{algorithm}[htb]
	\caption{TIA Portal logic data extraction pipeline}~\label{alg:tia_extraction_language}
	\footnotesize
	\begin{algorithmic}[1]
		\renewcommand{\algorithmicrequire}{\textbf{Input:}}
		\renewcommand{\algorithmicensure}{\textbf{Output:}}

		\REQUIRE XML data \(X\)
		\ENSURE Extracted logic blocks \(B\)

		\vspace{0.5em}
		\textit{initialize} :
		\vspace{0.2em}
		\STATE \(B \gets \emptyset\)

		\vspace{0.2em}
		\textit{iterate logic blocks}
		\vspace{0.2em}
		\FOR{each block \(b \in X.findall(\text{SW.Blocks.*})\)}
		\STATE \(language \gets \text{DetermineLanguage}(b)\)

		\vspace{0.2em}
		\textit{choose parsing strategy}
		\vspace{0.2em}
		\IF{\(language = \text{LAD} \lor language = \text{FBD}\)}
		\STATE \(parsed \gets \text{ExtractGraphicalLanguage}(b)\)
		\ELSIF{\(language = \text{SCL}\)}
		\STATE \(parsed \gets \text{ExtractSCLCode}(b)\)
		\ELSIF{\(language = \text{STL}\)}
		\STATE \(parsed \gets \text{ExtractSTLCode}(b)\)
		\ENDIF
		\STATE \(B.\text{append}(parsed)\)
		\ENDFOR

		\STATE \textbf{return} \(B\)
	\end{algorithmic}
\end{algorithm}

The extraction has three relevant limitations. First, comments and free-text annotations
are not consistently available across projects and are therefore only used when present.
Second, vendor specific blocks are captured syntactically but may not be fully accessible
and normalized semantically. Third, interpretation quality for signal intent can still
depend on naming quality and project-specific conventions. The goal of the extraction is
not to reproduce the full graphical layout of TIA Portal networks, but to preserve the
subset of information that is required for querying and reasoning about program behavior.

\FloatBarrier%
\subsection{Kinematic Model Extraction from Siemens NX MCD}

This subsection deterministically extracts kinematic metadata to prepare graph structured
input. Kinematic data is accessed from the Siemens NX MCD model of the manufacturing
system through the NX Open API at assembly level. We therefore implemented a dedicated
extraction pipeline for Siemens NX MCD to transform the proprietary, object-oriented
representation into a semantically structured intermediate format suitable for knowledge
graph integration.

The extraction process targets components that are relevant for VCM creation and
cross-domain reasoning. These include structural components and ownership relations and,
where available, rigid bodies, joints, actuators, and sensors. Together, these elements
represent the physical organization and motion-enabling constraints of the manufacturing
system and define the simulation objects to be linked to PLC signals and control logic.

The pipeline follows six steps: (1) load the NX MCD assembly, (2) traverse the component
hierarchy from root assemblies to leaf elements, (3) register category-specific observers
in an observer registry, (4) extract metadata per category during traversal, (5) preserve
ownership and structural relations between extracted objects, and (6) serialize the
resulting metadata into JSON for graph integration.

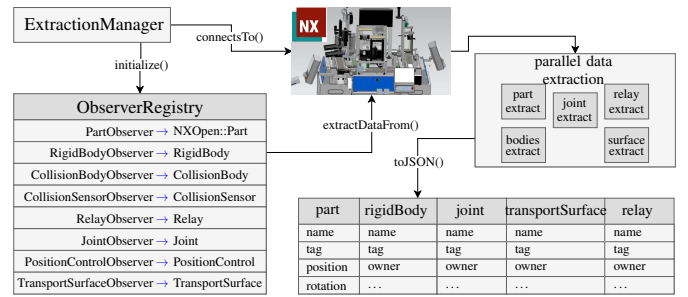
\begin{figure}[htb]
	\centering
	\resizebox{\columnwidth}{!}{%
		\begin{tikzpicture}
			\node[anchor=south west,inner sep=0] (image) at (0,0) {\includegraphics[width=\textwidth]{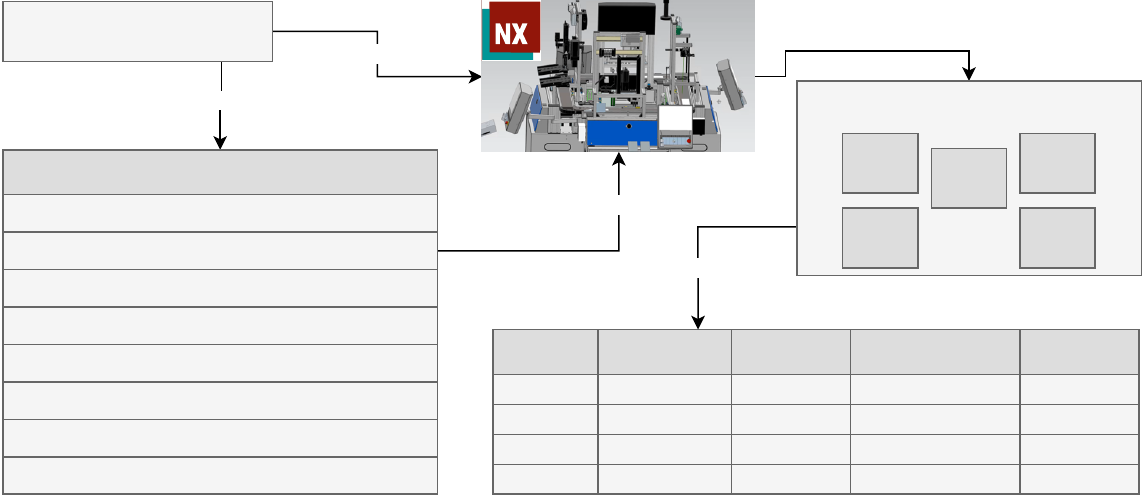}};
			\begin{scope}[x={(image.south east)},y={(image.north west)}]
				\Large
				\draw (.01,.933) node[anchor=west] {ExtractionManager};
				\draw (.193,.646) node[anchor=center] {ObserverRegistry};

				\normalsize
				\draw (.328,.89) node[anchor=center] {connectsTo()};
				\draw (.1953,.796) node[anchor=center] {initialize()};
				\draw (.54,.587) node[anchor=center] {extractDataFrom()};
				\draw (.612,.459) node[anchor=center] {toJSON()};

				\normalsize
				\draw (0.22,.565) node[anchor=east] {PartObserver};
				\draw (.21,.565) node[anchor=west] {\textcolor{blue}{$\rightarrow$} NXOpen::Part};
				\draw (0.22,.49) node[anchor=east] {RigidBodyObserver};
				\draw (.21,.49) node[anchor=west] {\textcolor{blue}{$\rightarrow$} RigidBody};
				\draw (0.22,.413) node[anchor=east] {CollisionBodyObserver};
				\draw (.21,.413) node[anchor=west] {\textcolor{blue}{$\rightarrow$} CollisionBody};
				\draw (0.22,.34) node[anchor=east] {CollisionSensorObserver};
				\draw (.21,.34) node[anchor=west] {\textcolor{blue}{$\rightarrow$} CollisionSensor};
				\draw (0.22,.261) node[anchor=east] {RelayObserver};
				\draw (.21,.261) node[anchor=west] {\textcolor{blue}{$\rightarrow$} Relay};
				\draw (0.22,.19) node[anchor=east] {JointObserver};
				\draw (.21,.19) node[anchor=west] {\textcolor{blue}{$\rightarrow$} Joint};
				\draw (0.22,.115) node[anchor=east] {PositionControlObserver};
				\draw (.21,.115) node[anchor=west] {\textcolor{blue}{$\rightarrow$} PositionControl};
				\draw (0.22,.038) node[anchor=east] {TransportSurfaceObserver};
				\draw (.21,.038) node[anchor=west] {\textcolor{blue}{$\rightarrow$} TransportSurface};
				\draw (.7685,.672) node[anchor=center] {\shortstack{part\\extract}};
				\draw (.7685,.52) node[anchor=center] {\shortstack{bodies\\extract}};
				\draw (.846,.641) node[anchor=center] {\shortstack{joint\\extract}};
				\draw (.924,.672) node[anchor=center] {\shortstack{relay\\extract}};
				\draw (.924,.52) node[anchor=center] {\shortstack{surface\\extract}};

				\large
				\draw (.8443,.78) node[anchor=center] {\shortstack{parallel data\\extraction}};
				\draw (.475,.2893) node[anchor=center] {part};
				\draw (.578,.2893) node[anchor=center] {rigidBody};
				\draw (.689,.2893) node[anchor=center] {joint};
				\draw (.818,.2893) node[anchor=center] {transportSurface};
				\draw (.9404,.2893) node[anchor=center] {relay};

				\normalsize
				\draw (.436,.2157) node[anchor=west] {name};
				\draw (.436,.156) node[anchor=west] {tag};
				\draw (.436,.0957) node[anchor=west] {position};
				\draw (.436,.0342) node[anchor=west] {rotation};

				\draw (.5288,.2157) node[anchor=west] {name};
				\draw (.5288,.156) node[anchor=west] {tag};
				\draw (.5288,.0957) node[anchor=west] {owner};
				\draw (.5288,.0268) node[anchor=west] {\dots};

				\draw (.6455,.2157) node[anchor=west] {name};
				\draw (.6455,.156) node[anchor=west] {tag};
				\draw (.6455,.0957) node[anchor=west] {owner};
				\draw (.6455,.0268) node[anchor=west] {\dots};

				\draw (.7495,.2157) node[anchor=west] {name};
				\draw (.7495,.156) node[anchor=west] {tag};
				\draw (.7495,.0957) node[anchor=west] {owner};
				\draw (.7495,.0268) node[anchor=west] {\dots};

				\draw (.898,.2157) node[anchor=west] {name};
				\draw (.898,.156) node[anchor=west] {tag};
				\draw (.898,.0957) node[anchor=west] {owner};
				\draw (.898,.0268) node[anchor=west] {\dots};

			\end{scope}
		\end{tikzpicture} }
	\caption[NX MCD data extraction pipeline architecture]{ Architectural overview of the NX
		MCD data extraction pipeline, illustrating the component-category observer that executes
		parallel extraction of key attributes from component classes into a JSON-format.}~\label{fig:mcd_extraction_pipeline}
\end{figure}

As illustrated in~\autoref{fig:mcd_extraction_pipeline}, the observer-based extraction architecture selects
category-specific metadata. Our architecture is open to extensibility: new NX MCD
component types can be integrated by adding dedicated observers without modifying the
traversal logic.

As in the PLC extraction pipeline, this step preserves task-relevant semantics rather
than reconstructing the full internal state of the authoring environment. It captures
only the kinematic entities and relations required for the downstream VCM tasks evaluated
in this study. Highly tool-specific configuration details that do not contribute to these
tasks are intentionally abstracted.

\FloatBarrier%
\subsection{Knowledge Graph Creation and Schema}~\label{subsec:graph_creation} This subsection deterministically enables joint retrieval
across both engineering domains, then extracted metadata are transformed into two
knowledge graphs in one database.

We implemented the knowledge graphs in Neo4j Graph Database~\cite{neo4j_neo4j_2026} because TIA and NX MCD data are heterogeneous, hierarchical, and
highly relational. Its property-graph model natively captures typed entities, relations,
and attributes. Neo4j property graph data model enables multi-hop Cypher execution to
utilize graphRAG~\cite{neo4j_neo4j_2026}.

Formally, the graphs are implemented as a labelled property-graph structure~\cite{skavantzos_normalizing_2023}
\[
	\mathcal{G}= (V, E, \Phi, \Psi, \lambda),
\]
where \(V\) denotes the finite set of entities (nodes), \(E\) the finite set of entities (edges),
\(\Phi: E \rightarrow V \times V\) maps each edge to its incident node pair,
\(\Psi: V \cup E \rightarrow \mathcal{P}(\mathcal{L})\) assigns labels to nodes and
edges, and \(\lambda:(V \cup E)\times \mathcal{K}\rightarrow\mathcal{N}\) assigns
property values to graph objects. Here, \(\mathcal{L}\) is the finite set of labels,
\(\mathcal{K}\) the set of property keys, and \(\mathcal{N}\) the set of property values. The overall node set is defined as the union of the
automation and kinematic domain entities,
\[
	V_{\mathrm{sys}} = V_{\mathrm{aut}} \cup V_{\mathrm{kin}},
\]
with
\[
	V_{\mathrm{aut}} = V_{\mathrm{variables}} \cup V_{\mathrm{blocks}} \cup V_{\mathrm{networks}},
\]
and
\[
	V_{\mathrm{kin}} = V_{\mathrm{parts}} \cup V_{\mathrm{bodies}} \cup V_{\mathrm{sensors}} \cup V_{\mathrm{joints}} \cup V_{\mathrm{actuators}}.
\]
The edge set is defined analogously as
\[
	E_{\mathrm{sys}} = E_{\mathrm{aut}} \cup E_{\mathrm{kin}},
\]
where \(E_{\mathrm{aut}}\) captures automation logic dependencies such as variable usage and
block calls, and \(E_{\mathrm{kin}}\) captures structural and simulation
relations such as hierarchy, joint assignment, and control attachment.

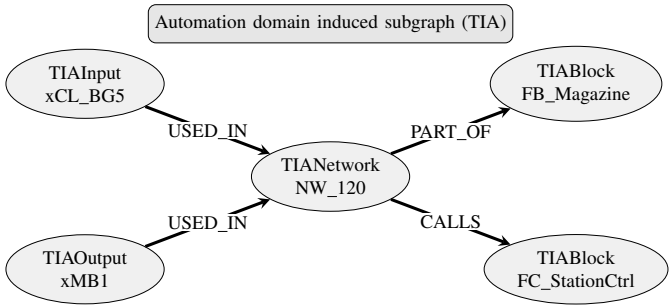
\begin{figure}[htb]
	\centering
	\resizebox{\columnwidth}{!}{%
		\begin{tikzpicture}[>=stealth]
			\tikzset{edge/.style={->, ultra thick}}
			\tikzset{ellip/.style={draw, ellipse, fill=black!6, align=center, minimum width=2.9cm,
						minimum height=1.05cm}}

			\node[draw, rounded corners, fill=black!10, minimum width=4.6cm, minimum height=0.7cm] at
			(0,2.8) {Automation domain induced subgraph (TIA)};

			\node[ellip] (v1) at (-4.5,1.7) {TIAInput\\xCL\_BG5};
			\node[ellip] (v2) at (-4.5,-1.7) {TIAOutput\\xMB1};
			\node[ellip] (n1) at (0,0.0) {TIANetwork\\NW\_120};
			\node[ellip] (b1) at (4.5,1.7) {TIABlock\\FB\_Magazine};
			\node[ellip] (b2) at (4.5,-1.7) {TIABlock\\FC\_StationCtrl};

			\draw[edge] (v1) -- node[pos=0.5, anchor=center, fill=white, inner sep=.8pt, outer
				sep=0pt] {USED\_IN} (n1);
			\draw[edge] (v2) -- node[pos=0.5, anchor=center, fill=white, inner sep=.8pt, outer
				sep=0pt] {USED\_IN} (n1);
			\draw[edge] (n1) -- node[pos=0.5, anchor=center, fill=white, inner sep=.8pt, outer
				sep=0pt] {PART\_OF} (b1);
			\draw[edge] (n1) -- node[pos=0.5, anchor=center, fill=white, inner sep=.8pt, outer
				sep=0pt] {CALLS} (b2);
		\end{tikzpicture} }
	\caption{Representative induced subgraph from the PLC domain showing typed node classes
		and edge types used for retrieval.}
	\label{fig:tia_subgraph}
\end{figure}

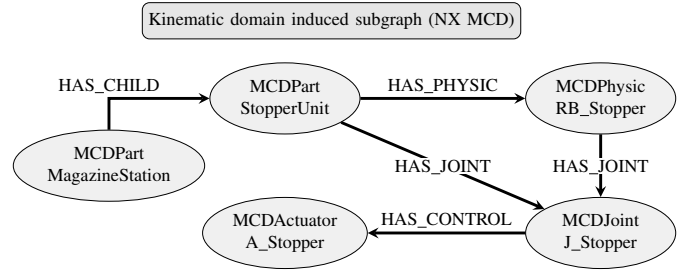
\begin{figure}[htb]
	\centering
	\resizebox{\columnwidth}{!}{%
		\begin{tikzpicture}[>=stealth]
			\tikzset{edge/.style={->, ultra thick}}
			\node[draw, rounded corners, fill=black!10, minimum width=5.2cm, minimum height=0.7cm] at
			(0.3,2.8) {Kinematic domain induced subgraph (NX MCD)};

			\node[draw, ellipse, fill=black!6, align=center, minimum width=2.9cm, minimum
				height=1.05cm] (p1) at (-4.0,0) {MCDPart\\MagazineStation};
			\node[draw, ellipse, fill=black!6, align=center, minimum width=2.9cm, minimum
				height=1.05cm] (p2) at (-0.6,1.3) {MCDPart\\StopperUnit};
			\node[draw, ellipse, fill=black!6, align=center, minimum width=2.9cm, minimum
				height=1.05cm] (rb1) at (5.5,1.3) {MCDPhysic\\RB\_Stopper};
			\node[draw, ellipse, fill=black!6, align=center, minimum width=2.9cm, minimum
				height=1.05cm] (j1) at (5.5,-1.3) {MCDJoint\\J\_Stopper};
			\node[draw, ellipse, fill=black!6, align=center, minimum width=2.9cm, minimum
				height=1.05cm] (a1) at (-0.6,-1.3) {MCDActuator\\A\_Stopper};

			\draw[edge] (p1.north) |- node[midway, above] {HAS\_CHILD} (p2.west);
			\draw[edge] (p2) -- node[midway, above] {HAS\_PHYSIC} (rb1);
			\draw[edge] (p2) -- node[pos=0.5,anchor=center,fill=white, inner sep=.8pt, outer sep=0pt]
			{HAS\_JOINT} (j1);
			\draw[edge] (rb1) -- node[pos=0.5,anchor=center,fill=white, inner sep=.8pt, outer sep=0pt]
			{HAS\_JOINT} (j1);
			\draw[edge] (j1) -- node[midway, above] {HAS\_CONTROL} (a1);
		\end{tikzpicture} }
	\caption{Representative induced subgraph from the NX MCD domain showing structural and
		control-relevant relation types used for retrieval.}
	\label{fig:mcd_subgraph}
\end{figure}

The graph construction follows three sequential steps. First, the normalized metadata
produced by the TIA and NX MCD extraction pipelines are transformed into graph entities
using deterministic mapping rules. Second, TIA attributes (such as data type, address,
network language, name and comment) and NX MCD attributes (such as tag, position,
rotation, speed, ownership, name and comment) are attached to the resulting nodes as
properties. Third, functional relations are instantiated explicitly as edges, see
\autoref{fig:tia_subgraph} and \autoref{fig:mcd_subgraph}.

Explicit cross-domain mappings between automation variables and NX MCD simulation objects
are not assumed to be fully known a priori but are instead addressed as part of the
downstream signal mapping task.

\FloatBarrier%
\subsection{Knowledge Retrieval and Agent Orchestration}~\label{subsec:agent_orch}

The multi-agent system was implemented using an open-source framework LangChain~\cite{langchain_langchain_2026} as orchestration layer and GPT-4o~\cite{openai_gpt-4_2023} as the underlying language model for all LLM-based agents.
LangChain was chosen as a pragmatic implementation framework because it provides modular
agent composition, tool integration, and controlled task routing~\cite{langchain_langchain_2026,holland_large_2024}. Other multi-agent frameworks are not
compared in this work.

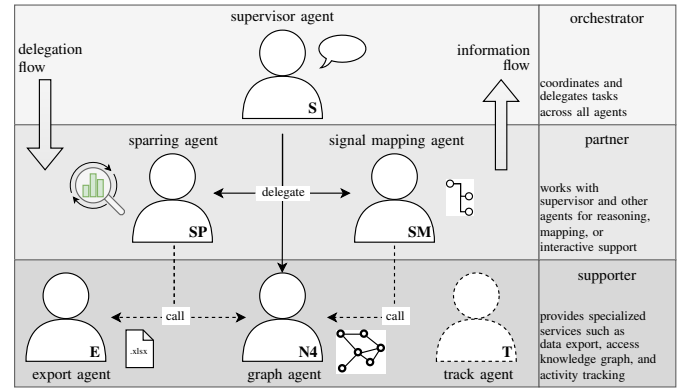
\begin{figure}[htbp]
	\centering
	\resizebox{\columnwidth}{!}{%
		\begin{tikzpicture}
			\node[anchor=south west,inner sep=0] (image) at (0,0) {\includegraphics[width=\textwidth]{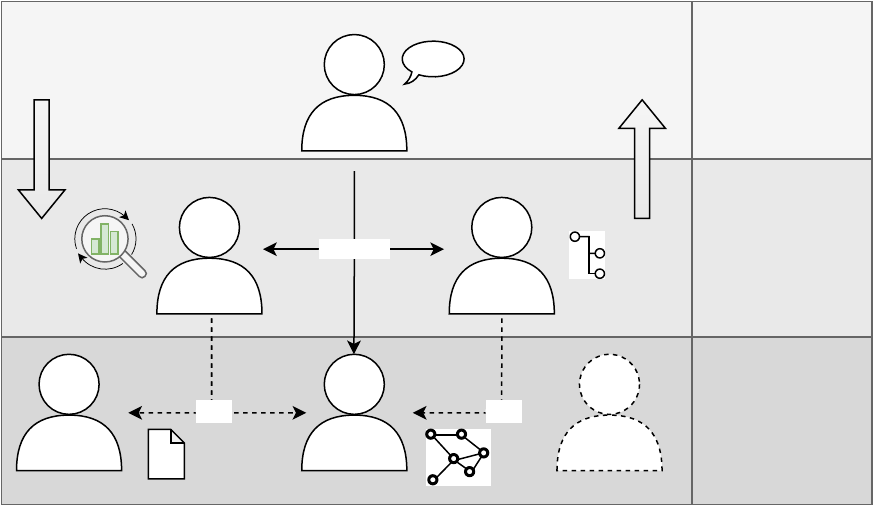}};
			\begin{scope}[x={(image.south east)},y={(image.north west)}]
				\large
				\draw (.405,.964) node[anchor=center] {supervisor agent};
				\draw (.241,.643) node[anchor=center] {sparring agent};
				\draw (.577,.643) node[anchor=center] {signal mapping agent};
				\draw (.088,.0327) node[anchor=center] {export agent};
				\draw (.408,.0327) node[anchor=center] {graph agent};
				\draw (.7,.0327) node[anchor=center] {track agent};

				\draw (.465,.701) node[anchor=south east] {\textbf{S}};
				\draw (.3,.378) node[anchor=south east] {\textbf{SP}};
				\draw (.637,.378) node[anchor=south east] {\textbf{SM}};
				\draw (.139,.0675) node[anchor=south east] {\textbf{E}};
				\draw (.466,.0675) node[anchor=south east] {\textbf{N4}};
				\draw (.759,.0675) node[anchor=south east] {\textbf{T}};

				\draw (.893,.964) node[anchor=center] {orchestrator};
				\draw (.893,.643) node[anchor=center] {partner};
				\draw (.893,.29) node[anchor=center] {supporter};

				\draw (.0,.86) node[anchor=west] {\shortstack[l]{delegation \\flow}};
				\draw (.7855,.86) node[anchor=east] {\shortstack[r]{information \\flow}};

				\normalsize
				\draw (.4065,.5083) node[anchor=center] {delegate};
				\draw (.244,.165) node[anchor=south] {call};
				\draw (.575,.165) node[anchor=south] {call};
				\draw (.7855,.69) node[anchor=south west] {\shortstack[l]{coordinates and\\delegates tasks\\across all agents}};
				\draw (.7855,.338) node[anchor=south west] {\shortstack[l]{works with\\supervisor and other\\agents for reasoning,\\mapping, or\\interactive support}};
				\draw (.7855,.006) node[anchor=south west] {\shortstack[l]{provides specialized\\services such as\\data export, access\\knowledge graph, and\\activity tracking}};
				\scriptsize
				\draw (.189,.0963) node[anchor=center] {.xlsx};
			\end{scope}
		\end{tikzpicture} }
	\caption[Proposed Multi-agent system architecture]{ Agent roles and interaction pathways
		in the multi-agent system, illustrating orchestration, collaboration, and specialist
		support among agents. Arrows indicate task and data flows.}~\label{fig:mas_architecture}
\end{figure}

Our role decomposition follows a hierarchical multi-agent coordination pattern known from
manufacturing multi-agent systems~\cite{andreadis_classification_2014}, where a central coordinator interacts with
specialized operational agents and additional supporting roles. We propose a three-tier
architecture, see~\autoref{fig:mas_architecture}, with separation into orchestrator, partner, and supporter
tiers. The orchestrator tier contains the supervisor agent for task classification and
routing. The partner tier contains the sparring agent and the signal mapping agent, which
transform retrieved engineering context into task-specific outputs. The supporter tier
contains the graph agent, export agent, and track agent, which provide graph retrieval,
output serialization, and runtime logging.

Delegation is top-down following insights from~\cite{andreadis_classification_2014}, while results and execution status propagate upward.
Agent memory is scoped to the current request context; no persistent cross-request memory
is used in this study.

\begin{table}[htb]
	\centering
	\caption{Main framework components and their task-specific interfaces.}
	\begin{tabular}{p{0.195\linewidth}|p{0.248\linewidth}|p{0.41\linewidth}}
		\toprule \textbf{Component} & \textbf{Input}      & \textbf{Output}                \\
		\midrule PLC extraction     & TIA Portal XML      & JSON (metadata)                \\[-0.2em]
		                            & export              &                                \\
		NX MCD ex-                  & NX MCD assem-       & JSON (simulation components)   \\[-0.2em]
		traction                    & bly                 &                                \\
		graph transfor-             & PLC and NX MCD      & graph nodes, properties, and   \\[-0.2em]
		mation                      & JSON                & edges                          \\
		supervisor                  & user request        & delegated task description     \\[-0.2em]
		agent                       &                     &                                \\
		graph agent                 & task description    & structured graph context       \\[-0.2em]
		                            & and graph schema    &                                \\
		sparring agent              & graph context       & NX Open journal script         \\
		signal mapping              & graph context (both & ranked mapping table \& NX     \\[-0.2em]
		agent                       & domains)            & Open journal                   \\
		export agent                & structured result   & text response and table        \\[-0.2em]
		                            & objects             &                                \\
		track agent                 & agent execution     & token usage and execution time \\[-0.2em]
		                            & state               & logs                           \\
		\bottomrule
	\end{tabular}\label{tab:components}
\end{table}

The selected LLM settings were determined empirically during development to obtain stable
behavior for the evaluated engineering tasks. \autoref{tab:components} summarizes the task-specific inputs and outputs of all framework
components. The parameters of the selected LLM are reported in~\autoref{tab:llm_parameter} for transparency.

\begin{table}[htb]
	\centering
	\caption{LLM-related configuration of the implemented agent workflow.}
	\begin{tabular}{l|c|c}
		\toprule \textbf{Agent}   & \textbf{Temperature} & \textbf{Max tokens} \\
		\midrule supervisor agent & 0.30                 & 4096                \\
		sparring agent            & 0.05                 & 4096                \\
		signal mapping agent      & 0.20                 & 4096                \\
		graph agent               & 0.10                 & 4096                \\
		export agent              & 0.15                 & 4096                \\
		track agent               & no LLM               & --                  \\
		\bottomrule
	\end{tabular}\label{tab:llm_parameter}
\end{table}

For knowledge-grounded tasks, the supervisor forwards a structured request to the graph
agent, which generates a Cypher query (a declarative graph query language for Neo4j)
using knowledge graph schema and task constraints. The retrieval process follows a
three-step strategy: initial LLM-query generation based on task context, deterministic
correction of syntactic or schema-related errors, and query refinement in case of empty
or insufficient results. Retrieved context from the graph database is returned in
structured form and passed to downstream agents. GraphRAG is used instead of vector-only
retrieval because the knowledge is represented as structured entities and typed
relations. The evaluated tasks require deterministic access to block-call,
variable-usage, and ownership relations through schema-constrained graph queries rather
than retrieval based primarily on semantic similarity.

The signal mapping agent uses retrieved graph context and returns ranked candidate lists
of NX MCD objects for a given PLC variable using lexical, type-based, and structural
cues. The export agent serializes output into tabular formats, while the track agent
records execution time and token usage.
\FloatBarrier%
\subsection{Evaluation Protocol}

This evaluation covers the three task classes introduced in~\autoref{sec:introduction}. Output quality was assessed for correctness, completeness, and
grounding. Grounding required that each output claim be supported by the retrieved graph
context and contain no unsupported technical facts, as verified by a domain expert
familiar with the laboratory manufacturing system. Table~\ref{tab:eval_queries} lists the evaluation set query used for the three use case groups.

\begin{table}[htbp]
	\caption{Evaluation query set used in the protocol across UC1--UC3.}
	\centering
	\begin{tabular}{c|p{0.387\textwidth}}
		\toprule \textbf{Acro} & \textbf{Questions}                                                                                        \\
		\midrule UC1.Q1        & Which component detects if the magazine is empty?                                                         \\
		UC1.Q2                 & What variable activates the conveyor belt motor?                                                          \\
		UC1.Q3                 & What controls the cylinder of the stopper?                                                                \\
		UC1.Q4                 & What TIA variable enables inch-travel mode of the DC motor?                                               \\
		UC1.Q5                 & Which outputs command separator close and separator open, and in which networks
		are those outputs written?                                                                                                         \\
		UC1.Q6                 & Which networks use xCL\_BG7 (pallet available) and xCL\_BG8 (front cover available),
		and where are both variables evaluated?                                                                                            \\
		UC1.Q7                 & What does it mean, when TIA Variable \%Q1.1 is TRUE?                                                      \\
		UC1.Q8                 & List all Festoplant MCD actuators, the joint each controls, and their
		position/speed properties, then return a table.                                                                                    \\
		UC1.Q9                 & Rank all variables (TIA and MCD) by distinct usage, return name, ID, domain,
		block calls, and networks usage, then export to Excel.                                                                             \\
		\midrule UC2.Q1        & Generate a NX MCD component for the mapped sensor corresponding to CL\_BG5.                               \\
		UC2.Q2                 & Generate a NX MCD component for the rigid body RB\_Stopper as a linear lifting cylinder along the Z-axis. \\
		UC2.Q3                 & Find the TIA component indicating the magazine lifting cylinder is in the upper
		position, then generate the corresponding NX MCD component from it.                                                                \\
		\midrule UC3.Q1        & Map the TIA Variable xCL\_BG5 to an NX MCD sensor.                                                        \\
		UC3.Q2                 & Map the TIA Variable xMB1 to an NX MCD actuator.                                                          \\
		UC3.Q3                 & Map TIA variable xCL\_BG1 to the most likely NX MCD component.                                            \\
		UC3.Q4                 & Map TIA variables xMB1 and xMB2 each to an NX MCD actuator.                                               \\
		\bottomrule
	\end{tabular}
	\label{tab:eval_queries}
\end{table}

For system-understanding queries, these criteria were applied directly to the returned
answers. NX Open journal generation was evaluated by successful execution of the
generated script in NX MCD without manual correction. Signal mapping was evaluated by the
plausibility and consistency of the proposed correspondences between PLC variables and
simulation objects and, where no unique mapping could be assumed a priori, by whether the
correct target appeared among the top-ranked suggestions.

Each use case was executed ten times to assess stability across runs. Where objective
checks were available, correctness was validated deterministically; otherwise, it was
assessed by expert review. In this study, token usage is treated as a proxy for LLM
computational demand and context complexity rather than as a direct quality metric. The
resulting measurements were compared across task classes with respect to computational
cost, robustness, and output quality, as reported in~\autoref{sec:results}.

A limitation of this approach is that output quality depends on the completeness of the
retrieved graph context and the coverage of supported output templates. The framework
therefore supports semi-automated VCM workflows but does not replace final expert
validation.

\section{Results}~\label{sec:results} This section provides quantitative and qualitative results for the
evaluated knowledge-graph-grounded multi-agent framework. We baselined our framework
against LLM-only, single-agent and human-evaluator approaches. As context for the
task-specific results,~\autoref{tab:elements_count} summarizes the size and composition of the extracted source
elements used in the evaluation. In total, the extraction yielded 52,268 automation
elements and 3,404 kinematic simulation elements.

\begin{table}[htb]
	\centering
	\caption{Extracted source elements used for evaluation.}
	\begin{tabular}{l|l|c}
		\toprule \textbf{Domain} & \textbf{Element category}       & \textbf{Count} \\
		\midrule TIA Portal      & variables                       & 2,511          \\
		TIA Portal               & PLC blocks                      & 399            \\
		TIA Portal               & networks                        & 2,012          \\
		TIA Portal               & network access elements         & 13,788         \\
		TIA Portal               & network parts                   & 8,983          \\
		TIA Portal               & wire connections                & 24,894         \\
		TIA Portal               & block call references           & 266            \\
		NX MCD                   & parts                           & 3,073          \\
		NX MCD                   & physical simulation components  & 248            \\
		NX MCD                   & sensing and logic components    & 23             \\
		NX MCD                   & joint components                & 32             \\
		NX MCD                   & actuation and motion components & 28             \\
		\bottomrule
	\end{tabular}\label{tab:elements_count}
\end{table}

\FloatBarrier%

\subsection{System Understanding}

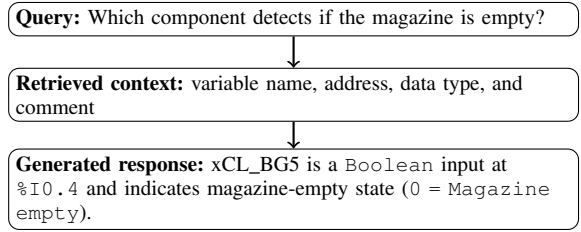
\begin{figure}[htb]
	\centering
	\footnotesize
	\begin{tikzpicture}
		\node[draw, rounded corners, align=left, text width=0.83\linewidth] (q) at (0,0) {\textbf{Query:} Which component detects if the magazine is empty?};
		\node[draw, rounded corners, align=left, text width=0.83\linewidth] (c) at (0,-1) {\textbf{Retrieved context:} variable name, address, data type, and comment};
		\node[draw, rounded corners, align=left, text width=0.83\linewidth] (r) at (0,-2.25) {\textbf{Generated response:} xCL\_BG5 is a \texttt{Boolean} input at
			\texttt{\%I0.4} and indicates magazine-empty state (\texttt{0 = Magazine empty}).};
		\draw[->, thick] (q) -- (c);
		\draw[->, thick] (c) -- (r);
	\end{tikzpicture}
	\caption{Compact example of a system-understanding query workflow from user question to
		grounded response.}
	\label{fig:UC1_example}
\end{figure}

This class evaluates the framework's ability to answer exploratory and explanatory
queries about the structure and behavior of the discrete manufacturing system based on
the overall knowledge representation. The main evaluation criteria were correctness,
grounding, and completeness with respect to the retrieved engineering context.

\autoref{tab:UC1_quant_res} summarizes the quantitative results for this task class. \autoref{fig:UC1_example} provides an illustrative example of a system-understanding query
and the corresponding response. Across ten runs per query, most UC1 responses were
correct and grounded.

\begin{table}[htb]
	\centering
	\caption{Performance of the system-understanding task across ten runs.}
	\begin{tabular}{l|c|c|c|c|c}
		\toprule \textbf{Query} & \textbf{Token} & \textbf{Time (s)} & \textbf{Correct} & \textbf{Grounded} & \textbf{Complete} \\
		\midrule UC1.Q1         & 3112           & 17.3              & 10/10            & 10/10             & 10/10             \\
		UC1.Q2                  & 3357           & 24.0              & 10/10            & 10/10             & 7/10              \\
		UC1.Q3                  & 2264           & 17.9              & 10/10            & 10/10             & 10/10             \\
		UC1.Q4                  & 3113           & 38.4              & 9/10             & 9/10              & 9/10              \\
		UC1.Q5                  & 2821           & 38.2              & 10/10            & 10/10             & 8/10              \\
		UC1.Q6                  & 2886           & 37.5              & 9/10             & 9/10              & 8/10              \\
		UC1.Q7                  & 2358           & 31.5              & 10/10            & 10/10             & 10/10             \\
		UC1.Q8                  & 1738           & 23.0              & 10/10            & 10/10             & 9/10              \\
		UC1.Q9                  & 11097          & 56.3              & 6/10             & 6/10              & 6/10              \\
		\bottomrule
	\end{tabular}\label{tab:UC1_quant_res}
\end{table}

Observed deficits were primarily related to completeness, with incomplete responses in
UC1.Q2 (3/10), UC1.Q5 (2/10), UC1.Q6 (2/10), and UC1.Q8 (1/10). In addition, UC1.Q4
produced one ungrounded response after failed graph retrieval, and UC1.Q9 showed four
graph-database memory overflows caused by a broad union query.

\FloatBarrier%

\subsection{Simulation Component Generation}
This task class evaluates the framework's ability to generate executable NX Open journal
scripts for extending or configuring the VCM in NX MCD. The primary evaluation criterion
was therefore successful script execution without manual correction for the right
component. For each query, the sparring agent retrieved the required component context
from the graph agent and selected an appropriate template for the requested simulation
object. The template was then parameterized with graph-derived metadata such as object
identifiers, component types, and configuration values. \autoref{tab:UC2_quant_res} summarizes the quantitative results for this task class.

\begin{table}[htb]
	\centering
	\caption{Performance of the simulation component generation task across ten runs.}
	\begin{tabular}{l|c|c|c|c}
		\toprule \textbf{Query} & \textbf{Token} & \textbf{Time (s)} & \textbf{Executable} & \textbf{Manual correction} \\
		\midrule UC2.Q1         & 1728           & 41.9              & 10/10               & 0/10                       \\
		UC2.Q2                  & 1270           & 14.7              & 10/10               & 0/10                       \\
		UC2.Q3                  & 3626           & 104.5             & 10/10               & 0/10                       \\
		\bottomrule
	\end{tabular}\label{tab:UC2_quant_res}
\end{table}

UC2.Q1 and UC2.Q2 required fewer tokens and less runtime (according to \autoref{tab:UC2_quant_res}) than UC2.Q3 because their requests specified the target
component more explicitly. UC2.Q3 imposed an additional identification step before
template instantiation, increasing computational cost; nevertheless, all runs produced
executable NX Open journals without manual correction.

\FloatBarrier%
\subsection{Cross-Domain Signal Mapping}

This task class evaluates the framework's ability to identify correspondences between PLC
variables from TIA Portal and control-relevant objects in the NX MCD assembly. Unlike
system-understanding queries or template-based script generation, signal mapping requires
explicit reasoning across two heterogeneous engineering domains.

\autoref{tab:UC3_quant_res} summarizes the quantitative results for this task class. In
addition to execution time and token consumption, we report Hit@1 as the task-quality
metric and the agent's self-reported confidence as an auxiliary diagnostic value. The
predicted mappings are then compared to reference mappings defined manually by a domain
expert. Across UC3.Q1-UC3.Q4, the top-ranked mapping was correct in all runs. Confidence
represents the average of ten runs confidence of the mapped signal.
\begin{table}[htb]
	\centering
	\caption{Performance of the cross-domain signal mapping task across ten runs.}
	\begin{tabular}{l|c|c|c|c}
		\toprule \textbf{Query} & \textbf{Token} & \textbf{Time (s)} & \textbf{Hit@1} & \textbf{Confidence} \\
		\midrule UC3.Q1         & 3864           & 39.4              & 10/10          & 89\%                \\
		UC3.Q2                  & 4410           & 53.6              & 10/10          & 85\%                \\
		UC3.Q3                  & 6574           & 66.6              & 10/10          & 91\%                \\
		UC3.Q4                  & 4225           & 61.5              & 10/10          & 78\%                \\
		\bottomrule
	\end{tabular}\label{tab:UC3_quant_res}
\end{table}

\FloatBarrier%
\subsection{Baseline Comparison and Observed Limitations}

\autoref{tab:compact_baseline} shows a clear separation between the evaluated
configurations. The LLM-only baseline scored 0.0\%
across all task classes because, without access to the graph
database, it could not use domain-specific information
from Siemens TIA Portal and NX MCD. The knowledge-graph-grounded single-agent baseline improved performance in all three use cases, while the proposed framework
achieved the highest scores among the automated configurations, with an overall score of 94.8\%. The comparison
includes only task-quality scores; token consumption and
execution time are excluded. The UC score reported in~\autoref{tab:compact_baseline} was calculated as the mean task-quality score across all
evaluation queries of the respective use case. For UC1, this score corresponds to the
average of correctness, grounding, and completeness across all runs and queries; for UC2,
it is based on executable generation without manual correction; for UC3, it is based on
Hit@1.

\begin{table}[htb]
	\centering
	\caption{Compact comparison of baseline configurations and the proposed framework across
		UC1--UC3.}
	\begin{tabular}{p{0.22\linewidth}|C{0.145\linewidth}|C{0.145\linewidth}|C{0.145\linewidth}|C{0.1\linewidth}}
		\toprule \textbf{Configuration} & \textbf{UC1 score (\%)} & \textbf{UC2 score (\%)} & \textbf{UC3 score (\%)} & \textbf{Overall (\%)} \\
		\midrule LLM-only               & 0.0                     & 0.0                     & 0.0                     & 0.0                   \\
		graph-grounded                  & 55.5                    & 66.6                    & 25.0                    & 49.0                  \\[-0.2em]
		single-agent                    &                         &                         &                         &                       \\
		proposed frame-                 & 90.7                    & 100.0                   & 100.0                   & \textbf{94.8}         \\[-0.2em]
		work                            &                         &                         &                         &                       \\
		\bottomrule
	\end{tabular}
	\label{tab:compact_baseline}
\end{table}

Although the framework produced technically valid outputs across the evaluated task
classes, four operational limitations were observed. First, applicability depends on
graph coverage and source-data availability. The LLM-only baseline confirmed that,
without graph access and engineering data, no usable outputs are produced. Likewise, if
one domain is missing from the graph (e.g., NX MCD), corresponding cross-domain tasks
such as signal mapping cannot be solved. Second, output quality depends on source-data
completeness and naming quality. Missing semantic information, incomplete metadata, and
weak naming conventions reduced retrieval quality and increased ambiguity during
cross-domain alignment. Third, graph-query scalability is limited for very broad
union-style requests over highly interconnected PLC structures. This behavior was
observed for UC1.Q9, where a single query can exceed available database memory; reliable
execution required decomposition into narrower task-focused queries. Fourth, component
generation is constrained by template coverage. Unsupported NX MCD component classes
cannot be generated automatically and therefore remain outside the current scope. These
limitations define the current operating boundaries of the framework rather than
invalidating its technical feasibility.

\FloatBarrier%

\section{Conclusion}~\label{sec:conclusion} This paper presented a knowledge-graph-grounded multi-agent
framework for semi-automated support of VCM development in discrete manufacturing. The
framework addresses a recurring practical challenge: engineering knowledge required for
VCM development is fragmented between PLC engineering and the mechanical domain where
early-stage data is often limited to static 3D CAD geometry. Integrating this information
still demands substantial manual effort and cross-domain expertise.

The proposed approach integrates structured data extracted from Siemens TIA Portal and
Siemens NX MCD into a graph-based knowledge representation and employs a hierarchical
multi-agent architecture to support three task classes: system understanding, simulation
component generation, and cross-domain signal mapping. Evaluated on a laboratory-scale
discrete manufacturing system, the framework showed consistency across all three task
classes. System-understanding queries were answered correctly and with full grounding in
most runs, all simulation component generation queries produced executable NX Open
journal scripts without manual correction, and cross-domain signal mapping returned the
correct top-ranked result in all runs. The baseline comparison further showed that the
proposed framework outperformed both the LLM-only and the graph-grounded single-agent
baselines.

The more relevant question is not which agent architecture performs best in isolation,
but how efficiently a graph-grounded framework can support engineers across tasks that
currently require manual cross-domain interpretation. This work contributes that
structured extraction, graph-based retrieval, and agent-based task execution can be
integrated into a unified assistance workflow. The resulting workflow makes cross-domain
engineering knowledge accessible and actionable in early-stage VCM development while
retaining human-in-the-loop validation as a required final step.

However, operational limitations define the current scope. Output quality depends on
graph coverage and source-data completeness. Broad union-style queries can exceed
graph-database memory limits. Component generation is constrained by the coverage of the
implemented NX Open template library. The framework was validated on a single
laboratory-scale system, and transferability to other machine architectures and
toolchains remains to be demonstrated.

Future work should focus on bringing other engineering data into graph-ready form, extend
the approach to larger industrial cases, and more robust retrieval workflows.

\bibliographystyle{IEEEtran}
\bibliography{bibo}

\end{document}